\documentclass[twocolumn,showpacs,preprintnumbers,amsmath,amssymb]{revtex4}

\usepackage{graphicx}
\usepackage{dcolumn}
\usepackage{bm}
\usepackage{epsfig}
\usepackage{epsfig}
\usepackage{amsmath}
\usepackage{amsfonts}

\newcommand{\ra}{\rangle}
\newcommand{\ba}{\begin{eqnarray}}
\newcommand{\ea}{\end{eqnarray}}
\newcommand{\be}{\begin{equation}}
\newcommand{\ee}{\end{equation}}
\newcommand{\bea}{\begin{eqnarray}}
\newcommand{\eea}{\end{eqnarray}}
\newcommand{\ud}{\mathrm{d}}
\usepackage{theorem}
\newtheorem{theorem}{Theorem}

\newtheorem{definition}{Definition}

\newtheorem{proposition}{Proposition}

\theorembodyfont{\rmfamily}
\newtheorem{remark}{Remark}
\theoremstyle{break}

\def\QED{~\rule[-1pt]{5pt}{5pt}\par\medskip}

\def\n{\noindent}

\def\Re {\mathbb{R}}

\begin{document}


\title{Time Optimal Control of Coupled Qubits Under Non-Stationary
Interactions}

\author{Haidong Yuan}

  \email{hyuan@fas.harvard.edu}
\author{Navin Khaneja}%
 \email{navin@hrl.harvard.edu}
\affiliation{ Division of Engineering and Applied Science, Harvard
University, 33 Oxford Street, Cambridge MA 02138 }

\date{\today}

\begin{abstract}
In this article, we give a characterization of all the
unitary transformations that can be synthesized in a given time
for a two-qubit system in the presence of general time varying
coupling tensor. This characterization helps to compute the
minimum time and the shortest pulse sequence for generating a
general two-qubit transformation under non-stationary interactions.
The methods presented here can be applied in design of time optimal
pulse sequences for transferring coherence and polarization between coupled
spins with time varying couplings as in solid state NMR under
magic angle spinning.
\end{abstract}

\maketitle

\section{\label{sec:introduction}Introduction}
An important question in quantum information science is to
determine the minimum time required to perform a quantum
computation using a set of physical resources. Since two-qubit
gates are the building blocks of quantum information processing,
it is of fundamental interest to find the minimal time required to
implement a unitary operation on a two-qubit system using the
interaction Hamiltonian $H_d$, and the local unitary operations on
the two qubits. This problem was studied in~\cite{Khaneja:01},
where it was shown that any two qubit unitary propogator $U_F$ can
be expressed as

\begin{equation}\label{eq:topt}
U_F = U_2 \ (\prod_{k=1}^{4} \ V_k \exp(-i H_d
t_k)\ V_k^{\dagger})\ U_1,
\end{equation} where $U_1, U_2, V_k$ are local unitary
transformations and the effective Hamiltonians
$V_kH_dV_k^{\dagger}$ all mutually commute. Under the assumption
that the synthesis of local unitaries takes arbitrarily small
time, the minimum time to produce a desired $U_F$ is the smallest
value of $\sum_{k=1}^4 t_k$ in equation
(\ref{eq:topt})~\cite{Khaneja:01}. This characterization of time
optimal trajectories is used in ~\cite{Vidal} to explicitly
compute an elegant expression for the minimal time for synthesis
of arbitrary unitary transformation of two qubits. Alternate
proofs for time optimality have been presented in ~\cite{Hammerer,
Childs}. There is now a considerable literature on the subject;
see for example, ~\cite{Hammerer, Vidal, Bullok, Zhang,
Haselgrove, Childs, Zeier, Vatan, khanejanmr} and references
therein.

All these investigations assume that the interaction Hamiltonian
$H_d$ is fixed. In this paper, we consider the general problem
when $H_d$ varies with time. For example, in solid state
NMR~\cite{Spiess}, the interaction between the spins are varying
with time during magic angle spinning when the sample is rotated
around an axis making an angle of $\theta_M=\tan^{-1}(\sqrt{2})$
with the static magnetic field $B_0$. As a result the dipolar
couplings between nuclear spins that have an orientational
dependence of the form $3\cos^2(\theta)-1$ averages out( $\theta$
is the angle of internuclear axis with the static magnetic field),
leading to better resolved NMR spectrum~\cite{Spiess}. An
important problem in multi-dimensional solid state NMR experiments
is to find radio-frequency pulse sequence that re-couple desired
spins whose interactions are being modulated in time by magic
angle spinning. Finding short pulse sequences that transfer
polarization or coherence between coupled nuclear spins under time
varying interactions is of interest in solid state NMR. In this
paper, we give a complete characterization of all the unitary
transformations that can be synthesized in a given time for a
two-qubit system in presence of general time varying coupling
tensor, assuming that the local unitary transformation on two
qubits can be performed arbitrarily fast(on a time scale governed
by the strength of couplings). From the perspective of quantum
control theory, this problem is equivalent to characterizing the
reachable set of the Schr$\ddot{o}$dinger equation
\begin{equation}\label{eq:main}
\dot{U}(t)=-i[H_d(t)+\sum_{j=1}^{m}v_j(t)H_j]U(t),
\end{equation}
where $U\in SU(4)$ and $H_d(t)$ is the interaction Hamiltonian
that is internal to the system and $\sum_{j=1}^m v_j(t)H_j$ is the
part of the Hamiltonian that can be externally changed, and
generates the local unitary operations. We assume the control
parameters $v_j$ are a priori not bounded.

Before stating the main result, we review some background
material.

\section{BACKGROUND}
\subsection{Majorization}
For an element $x=(x_1 , . . . ,x_k)^T$ of $\Re^k$ we denote by
$x^\downarrow =(x_1^\downarrow , . . . ,x_k^\downarrow)^T$ a
permutation of $x$ so that $x_i^\downarrow\geq x_j^\downarrow$ if
$i<j$, where $1\leq i, j\leq k$.
\begin{definition}[majorization]{\rm
A vector $x\in \Re^k$ is majorized by a vector $y\in \Re^K$
(denoted $x \prec y$), if
\begin{equation}
  \sum_{j=1}^k x^{\downarrow}_j \leq \sum_{j=1}^k y^{\downarrow}_j
\end{equation}
for $k = 1,\ldots,D-1$, and the inequality holds with equality
when $k= D$.}
\end{definition}
\begin{proposition}
\label{prop:convexhull} {\rm $x \prec y$ iff $x$ lies in the convex
hull of $y$ and all its permutations $P_iy$, where $P_i$ are
permutation matrices.}
\end{proposition}
\begin{proposition}[Schur, Horn ~\cite{Bhatia,Horn}]
\label{prop:Schur} {\rm For an element $\lambda=(\lambda_1 , . . .
,\lambda_n)^T$, let $D_\lambda$ be a diagonal matrix with
$(\lambda_1 , . . . ,\lambda_n)$ as its diagonal entries, let
$a=(a_1,...,a_n)^T$ be the diagonal entries of matrix
$A=K^TD_\lambda K$, where $K\in SO(n)$. Then $a \prec \lambda$.
Conversely for any vector $a \prec \lambda$, there exists a $K\in
SO(n)$, such that $(a_1,...,a_n)^T$  are the diagonal entries of
$A=K^TD_\lambda K$}
\end{proposition}

Following \cite{Bennett,Hammerer,Vidal}, for an element $x=(x_1,x_2,x_3)^T$ of $\Re^3$,
we introduce the vector $\hat{x}=(|x_1|,|x_2|,|x_3|)^T$, and define the $s$-order
version $x^s$ of $x$ by setting $x_1^s=\hat{x}_1^\downarrow$,
$x_2^s=\hat{x}_2^\downarrow$,
$x_3^s=sgn(x_1x_2x_3)\hat{x}_3^\downarrow$.
\begin{definition}[\cite{Bennett,Hammerer,Vidal}] {\rm
The vector $x\in \Re^3$ is $s$-majorized by $y\in \Re^3$(denoted
$x\prec_s y$) if
\begin{eqnarray}
\label{eq:s-major}
\aligned
x_1^s&\leq y_1^s \\
x_1^s+x_2^s+x_3^s&\leq y_1^s+y_2^s+y_3^s\\
x_1^s+x_2^s-x_3^s&\leq y_1^s+y_2^s-y_3^s
\endaligned
\end{eqnarray}}
\end{definition}

\subsection{Canonical Decomposition}
 An arbitrary two-qubit Hamiltonian can be parameterized
\begin{equation}
H_d(t)=I\otimes(\vec{a}(t)\cdot\vec{\sigma}) +
(\vec{b}(t)\cdot\vec{\sigma})\otimes I + \sum_{i,j} M_{ij}(t)
\sigma_i \otimes \sigma_j
\end{equation}
where $i,j\in\{x,y,z\}$ and $\vec{a}\equiv(a_x,a_y,a_z)$,
$\vec{b}\equiv(b_x,b_y,b_z)$ are real 3-vectors, $M$ is a 3 by 3
real matrix, and $\vec{\sigma}=(\sigma_x,\sigma_y,\sigma_z)$ is
the vector of Pauli operators.

Let $H_d'(t)$ be the non-local part of $H_d(t)$, i.e.,
$$H_d'(t)=\sum_{i,j} M_{ij}(t) \sigma_i \otimes \sigma_j.$$
Since we assume that the local unitaries can be generated in
arbitrarily small time, all the unitaries transformations that can
be synthesized in a given time under $H_d(t)$ can also be
synthesized under $H_d'(t)$ and vice versa \cite{Khaneja:01}. We
therefore consider $H_d(t)$ and $H_d'(t)$ are interchangeable
resources under fast local unitaries. From now on we
assume $H_d(t)$ has only non-local terms.
\begin{proposition}[Canonical Decomposition~\cite{Khaneja:01,Kraus} ]
\label{prop:canonical} {\rm Any two-qubit non-local Hamiltonian $H$
can be written in the form
\begin{equation}
H=(A\otimes B)^\dagger(\theta_1^H\sigma_x\otimes
\sigma_x+\theta_2^H\sigma_y\otimes
\sigma_y+\theta_3^H\sigma_z\otimes \sigma_z)(A\otimes B)
\end{equation}
and any two-qubit unitary $U\in SU(4)$ may be written in the form
\begin{equation}
U=(A_1\otimes B_1)e^{-i(\theta_1^U\sigma_x\otimes
\sigma_x+\theta_2^U\sigma_y\otimes \sigma_y+\theta_3^UZ\otimes
Z)}(A_2\otimes B_2)
\end{equation}
here $A$, $A_1$, $A_2$, $B$, $B_1$, $B_2$ are single-qubit
unitaries, and
\begin{eqnarray}
\label{eq:s-order}
\aligned
 \theta_1^H&\geq \theta_2^H\geq |\theta_3^H|\\
 \frac{\pi}{4}\geq &\theta_1^U\geq \theta_2^U\geq |\theta_3^U|
\endaligned
\end{eqnarray}
We call $\theta_1^H\sigma_x\otimes
\sigma_x+\theta_2^H\sigma_y\otimes
\sigma_y+\theta_3^H\sigma_z\otimes \sigma_z$ and
$e^{-i(\theta_1^U\sigma_x\otimes
\sigma_x+\theta_2^U\sigma_y\otimes
\sigma_y+\theta_3^U\sigma_z\otimes \sigma_z)}$ the canonical form
of $H$ and $U$ respectively, and $\vec{\theta}^H$ and
$\vec{\theta}^U$ the canonical parameters of $H$ and $U$
respectively. For a 3-vector $\vec{\beta}$, we denote
$$H_{\vec{\beta}}=\beta_1\sigma_x\otimes \sigma_x+\beta_2\sigma_y\otimes \sigma_y+\beta_3
\sigma_z\otimes \sigma_z$$
$$U_{\vec{\beta}}=e^{-i(\beta_1\sigma_x\otimes \sigma_x+\beta_2\sigma_y\otimes \sigma_y+\beta_3
\sigma_z\otimes \sigma_z)}$$}
\end{proposition}

\subsection{Magic Basis}{\label{subsec:magic}}
The magic basis is a vector space basis for two-qubit pure states:
\begin{eqnarray}
 \frac{|00\ra+|11\ra}{\sqrt{2}}; & &
i\frac{|00\ra-|11\ra}{\sqrt{2}}; \nonumber \\
i\frac{|01\ra+|10\ra}{\sqrt{2}}; & &
 \frac{|01\ra-|10\ra}{\sqrt{2}}.
\end{eqnarray}
The basis change from the standard basis
$\{|00\ra,|01\ra,|10\ra,|11\ra\}$ to the magic basis is given by
$Q^{-1}$, where $$Q=\frac{1}{\sqrt{2}}\left(\begin{array}{cccc}
      1 & 0 & 0 & i\\
      0 & i & 1 & 0 \\
      0 & i & -1& 0 \\
      1 & 0 & 0 & -i \\

      \end{array}\right). $$

For elements $U\in SU(4)$ the map $U\rightarrow Q^{-1}UQ$ reflects
the isomorphism between $SU(2)\otimes SU(2)$ and $SO(4)$
\cite{Makhlin}. When expressed in the magic basis, canonical form
Hamiltonian and unitaries are diagonal. In magic basis, the
canonical decomposition takes the form $H_d=K^TD_HK$, $U=RD_US$,
where $K$, $R$ and $S$ are real orthogonal matrices, and $D_H$,
$D_U$ are diagonal matrices. The diagonal elements of $D_H$ and
$D_U$ are easily written in the terms of the canonical form
parameters $\theta^i,i\in\{H,U\}$. Define
\begin{eqnarray}
\label{eq:magic} \aligned
\varphi_1^i&=\theta_1^i+\theta_2^i-\theta_3^i,
\varphi_2^i=\theta_1^i-\theta_2^i+\theta_3^i,\\
\varphi_3^i&=-\theta_1^i+\theta_2^i+\theta_3^i,
\varphi_4^i=-\theta_1^i-\theta_2^i-\theta_3^i.
\endaligned
\end{eqnarray}
The diagonal elements of $D_H$ are $(\varphi_1^H,
\varphi_2^H, \varphi_3^H, \varphi_4^H)$ and the diagonal elements
of $D_U$ are
$(e^{-i\varphi_1^U},e^{-i\varphi_2^U},e^{-i\varphi_3^U},e^{-i\varphi_4^U})$.
Under choice of magic basis, $H_d$ is real symmetric and
$(\varphi_1^H, \varphi_2^H, \varphi_3^H, \varphi_4^H)$ are its
eigenvalues. Eq.(\ref{eq:s-order}) together with
Eq.(\ref{eq:magic}) implies that
$$\varphi_1^i\geq\varphi_2^i\geq\varphi_3^i\geq\varphi_4^i.$$

\begin{proposition}[\cite{Hammerer}]{\rm Let $\vec{\alpha}$ and $\vec{\beta}$ be two real
s-ordered 3-vectors, let $\vec{\lambda}$ and $\vec{\mu}$ be the 4-vectors
related to $\vec{\alpha}$ and $\vec{\beta}$ respectively via
(\ref{eq:magic}), then $\vec{\lambda}\prec \vec{\mu}$ iff
$\vec{\alpha} \prec_s \vec{\beta}$.}
\end{proposition}
The proof follows from the definitions.
\section{Result}
The main result of this paper is as follows:
\begin{theorem}{\rm
Let $\vec{\theta}^H(t)$ be the canonical parameters of $H_d(t)$
in (\ref{eq:main}) and $\vec{\theta}(T)=\int_{0}^{T}\theta^H(t) \ud t$, where the
integration is performed for each entry of the vector. All the
unitary operators that can be generated within time T with
$H_d(t)$ and fast local unitaries are given by the set
$${\cal{R}}(T) = \{K_1U_{\vec{\beta}}K_2| K_1,K_2\in SU(2)\otimes SU(2),\vec{\beta} \prec_s \vec{\theta}(T)\}$$}
\end{theorem}

\begin{remark}{\rm We prove this theorem by using the choice of magic basis. In this basis, $\{-iH_j\}$
are skew-symmetric matrices and
generate the group $SO(4)$. The interaction Hamiltonian $H_d$ can
be expressed as $H_d=K^TD_{\vec{\lambda}} K$, where $K\in SO(4)$
and $D_{\vec{\lambda}}$ a diagonal matrix with diagonal entry
$\vec{\lambda}$ related to $\vec{\theta}^H$ via (\ref{eq:magic}).
Let $\vec{\gamma}(T)=\int_{0}^{T}\vec{\lambda}(t) \ud t$, then
in the magic basis
$${\cal{R}}(T) = \{Re^{-iD_{\vec{\beta}}}S| R,S\in SO(4),\vec{\beta} \prec \vec{\gamma}(T)\}. $$ }
\end{remark}

\n {\bf Proof:} Under the choice of magic basis, we can write
$U(t)=R(t)A(t)S(t)$, where $R(t),S(t)\in SO(4)$ and $A(t)$ be
diagonal matrix. Assumption of fast local unitaries implies
we can generate $SO(4)$ instantly , so it
suffices to prove all we can generate for the $A$ part is
$e^{-iD_{\vec{\beta}}}$, $\vec{\beta} \prec \vec{\gamma}(T)$.

Assume $U(t)=R(t)A(t)S(t)$ is a trajectory of Eq.(\ref{eq:main}),
then $A(t)=R^T(t)U(t)S^T(t)$ and $\dot{A}(t)=$
$$\dot{R}^T(t)U(t)S^T(t)+R^T(t)\dot{U}(t)S^T(t) +R^T(t)U(t)\dot{S}^T(t). $$
Let $\dot{R}^T(t) = r(t) {R}^T(t)$ and $\dot{S}^T(t) =
{S}^T(t)s(t)$, substituting for $\dot{U}(t)$, we get
$R^T(t)\dot{U}(t)S^T(t)=$
$$ R^T(t)\ [-iH_d(t)\\
 -i\sum_{i=1}^{m}v_j(t) H_j ]R(t) ( R^T(t)U(t)S^T(t)). $$ Using $H_d(t) = K^T(t)D_{\vec{\lambda}} K(t)$
 we get $R^T(t)\dot{U}(t)S^T(t)=$ $$R^T(t)\ [-i K^T(t)D_{\vec{\lambda}} K(t)\\
 -i\sum_{i=1}^{m}v_j(t) H_j ]R(t) (A(t)). $$ Let $P(t) = K(t)R(t)$ and denote $h(t) =  R^T(t)\ [-i\sum_{i=1}^{m}v_j(t) H_j ]R(t)$. Equation for evolution of $A(t)$ then takes the form $\dot{A}(t)=$

\begin{equation} \label{eq:H}
r(t) A(t) + [P^T(t) D_{\vec{\lambda}}(t) P(t)]A(t) + h(t) A(t) + A(t) s(t) .
\end{equation}
Notice that $r(t),s(t)$, and $h(t)$ are in $so(4)$ (skew symmetric matrices of dimension 4) and hence their
diagonal entries are all zero. When
multiplied by a diagonal matrix $A(t)$, the diagonal entries
remain zero. Therefore in the evolution equation of $A(t)$,
these terms must sum to zero and we can discard these terms.
We get \begin{equation}
\label{eq:D} \dot{A}(t)=D_{-i\vec{\mu}(t)}A(t)
\end{equation}
where $\vec{\mu}(t)$ is the diagonal entries of $P^T(t)
D_{\vec{\lambda}(t)} P(t)$. Since we can generate elements of
$SO(4)$ in arbitrarily small time, $R(t)$ and hence $P(t)$ can
take value of any element in $SO(4)$ and from Proposition
(\ref{prop:Schur}), $\vec{\mu}(t)$ can take any element of the set
$\{\vec{\mu}(t)|\vec{\mu}(t)\prec \vec{\lambda}(t)\}$.

From Eq.(\ref{eq:D}), we get $A(T)=e^{-iD_{\vec{\beta}}}$, where
$\vec{\beta}=\int_{0}^{T}\vec{\mu}(t) \ud t$, $\vec{\mu}(t)\prec
\vec{\lambda}(t)$. We first prove $\vec{\beta}\prec
\vec{\gamma}(T)$, and then show that $\vec{\beta}$ can take on the
values of any vector majorized by $\vec{\gamma}(T)$.
\begin{equation}
\label{eq:gamma}
\sum_{j=1}^k\gamma^{\downarrow}_j(T)=\sum_{j=1}^k\int_{0}^{T}\lambda^{\downarrow}_j(t)\ud
t
\end{equation}
\begin{equation}
\label{eq:beta}
\sum_{j=1}^k\beta^{\downarrow}_j=\sum_{j=1}^k\int_{0}^{T}\mu_{\sigma(j)}(t)\ud
t
\end{equation}
where $\sigma$ is some permutations and $k=1,2,3,4$. On
subtracting Eq.(\ref{eq:beta}) from Eq.(\ref{eq:gamma}), we get
\begin{eqnarray}
\label{eq:Major}
 \aligned
\sum_{j=1}^k\gamma^{\downarrow}_j(T)-\sum_{j=1}^k\beta^{\downarrow}_j=
\int_{0}^{T}\sum_{j=1}^k\lambda^{\downarrow}_j(t)-\sum_{j=1}^k\mu_{\sigma(j)}(t)\ud
t.
\endaligned
\end{eqnarray}
Since $\mu(t)\prec \lambda(t)$,
$\sum_{j=1}^k\lambda^{\downarrow}_j(t)-\sum_{j=1}^k\mu_{\sigma(j)}(t)\geq
0$, and from Eq.(\ref{eq:Major}), $\sum_{j=1}^k\gamma^{\downarrow}_j(T)-\sum_{j=1}^k\beta^{\downarrow}_j\geq
0$. Obviously when $k=4$, both terms equal 0, the equality holds, so $\vec{\beta}\prec \vec{\gamma}(T)$.

We now prove $\vec{\beta}$ can take on the values of all the
vectors majorized by $\vec{\gamma}(T)$, which is the convex hull
of $\vec{\gamma}(T)$ and all its permutations. If we
take $R(t)=K^T(t)$, then $\vec{\beta}=\vec{\gamma}(T)$. It is
also easy to see $\vec{\beta}$ can take all the permutations of
$\vec{\gamma}(T)$, so we just need to prove that the vectors
$\vec{\beta}$ can reach is a convex set. Let $\alpha\in[0,1]$,
\begin{eqnarray}
\vec{\beta}_1=\int_{0}^{T}\vec{\mu}_1(t) \ud t\\
\vec{\beta}_2=\int_{0}^{T}\vec{\mu}_2(t) \ud t
\end{eqnarray}
then
\begin{eqnarray}
\alpha
\vec{\beta}_1+(1-\alpha)\vec{\beta}_1=\int_{0}^{T}\alpha\vec{\mu}_1(t)+(1-\alpha)\vec{\mu}_2(t)\ud
t
\end{eqnarray}
but $\alpha\vec{\mu}_1(t)+(1-\alpha)\vec{\mu}_2(t)\prec
\vec{\lambda}(t)$, so $\alpha
\vec{\beta}_1+(1-\alpha)\vec{\beta}_1$ can also be achieved.
\hfill{Q.E.D}\\

Given these theorems, we can compute the minimum time
needed to generate any unitary operator $U$ in $SU(4)$ with
$H_d(t)$ and fast local unitaries.
\begin{theorem}\label{thm:time}{\rm Using the Hamiltonian $H_d(t)$ and fast local unitaries, a
two-qubit gate $U$ can be generated within time T iff there exists
a vector $\vec{n}=(n_1,n_2,n_3)$ of integers, such that
$\vec{\beta}_{\vec{n}}=\vec{\theta}^{\ U} +\frac{\pi}{2}\vec{n}$
satisfies
$$\vec{\beta}_{\vec{n}}\prec_s
\int_{0}^{T}\vec{\theta}^{H_d(t)}\ud t$$ where $\vec{\theta}^{\ U}$
and $\vec{\theta}^{H_d(t)}$ are the canonical parameters of $U$
and $H_d(t)$ respectively. The minimum time required to simulate
$U$ is given by the minimum value of $T\geq 0$ such that either
\begin{equation}
\vec{\beta}_{(0,0,0)}\prec_s \int_{0}^{T}\vec{\theta}^{H_d(t)}\ud
t
\end{equation}
or
\begin{equation}
\vec{\beta}_{(-1,0,0)}\prec_s \int_{0}^{T}\vec{\theta}^{H_d(t)}\ud
t
\end{equation} holds}
\end{theorem}
The proof follows the treatment in ~\cite{Vidal}.

\n {\bf Proof:} Recall that all commutators $[\sigma_j\otimes
\sigma_j,\sigma_k\otimes \sigma_k]$ vanish, and that
$\exp(-i\frac{\pi}{2}\sigma_j\otimes \sigma_j)=-i\sigma_j\otimes
\sigma_j$ is a local gate. This implies that
$\vec{\theta}^{\ U} +\frac{\pi}{2}\vec{n}$ represents all vectors
compatible with the gate $U$. It is straightforward to check from
Eq.(\ref{eq:s-major}) that for any two vectors $\vec{x}$ and
$\vec{y}$, with components $x_1\geq x_2\geq |x_3|$, $y_1\geq
y_2\geq |y_3|$, if $y_1\geq 3x_1$, then $\vec{x}\prec_s\vec{y}$.
By definition $\frac{\pi}{4}\geq \theta^U_1\geq 0$, if some
component $n_j$ of $\vec{n}$ fulfills $|n_j|>1$, then the maximal
component of the reordered version of
$\vec{\theta}^{\ U} +\frac{\pi}{2}\vec{n}$ is at least
$\frac{3\pi}{4}$, which implies
$\vec{\theta}^{\ U} \prec_s\vec{\theta}^U+\frac{\pi}{2}\vec{n}$.
Therefore we can restrict our attention to vectors $\vec{n}$ with
$|n_j|\leq 1$. A case by case check shows that for $\vec{n}\in
\{(-1,-1,-1),(0,-1,0),(0,0,-1),(0,0,1)\}$,
$\vec{\theta}^{\ U} +\frac{\pi}{2}(-1,0,0)\prec_s\vec{\theta}^{\ U} +\frac{\pi}{2}\vec{n}$,
and for the remaining vectors $\vec{n}$,
$\vec{\theta}^{\ U} +\frac{\pi}{2}(0,0,0)\prec_s\vec{\theta}^{\ U} +\frac{\pi}{2}\vec{n}$.
Thus the result follows.

\section{Example}
We now work an explicit example on finding the minimum time to synthesize a
desired unitary under time varying couplings. Assume the interaction $H_d(t)$ takes the form
$D(t)(\sigma_x\otimes\sigma_x+\sigma_y\otimes\sigma_y-2\sigma_z\otimes\sigma_z)$.
We compute the minimum time to generate a swap gate corresponding to the unitary transformation
$U=\exp{-i\frac{\pi}{4}(\sigma_x\otimes\sigma_x+\sigma_y\otimes\sigma_y+\sigma_z\otimes\sigma_z)}$.

To fix ideas, consider the case when $D(t)$ is constant, say $D>0$. The
canonical parameters of $H_d(t)$ and $U$ are $D(2,1,-1)$ and
$\frac{\pi}{4}(1,1,1)$ respectively. The minimum time to generate
$U$ is the minimum $T$ that satisfies $\frac{\pi}{4}(1,1,1)\prec_s
DT(2,1,-1)$ or
$\frac{\pi}{4}(1,1,1)+\frac{\pi}{2}(-1,0,0)\prec_sDT(2,1,-1)$,
which is $\frac{3\pi}{16D}$. The strategy to generate $U$ is to use selective
excitation on first spin preparing an effective Hamiltonian
$D(-\sigma_x\otimes\sigma_x+\sigma_y\otimes\sigma_y+2\sigma_z\otimes\sigma_z)$,
which evolves $\frac{\pi}{16D}$ units of time. This is followed by evolution of effective Hamiltonians
$D(-2\sigma_x\otimes\sigma_x+\sigma_y\otimes\sigma_y+\sigma_z\otimes\sigma_z)$
and $D(-\sigma_x\otimes\sigma_x+2\sigma_y\otimes\sigma_y+\sigma_z\otimes\sigma_z)$
for $\frac{\pi}{16D}$ units of time each. In the end we apply a local unitaries
$e^{-i\frac{\pi}{2}\sigma_x\otimes\sigma_x}=-i\sigma_x\otimes\sigma_x$.

\begin{figure}[t]
\begin{center}

\includegraphics[width=1\hsize]{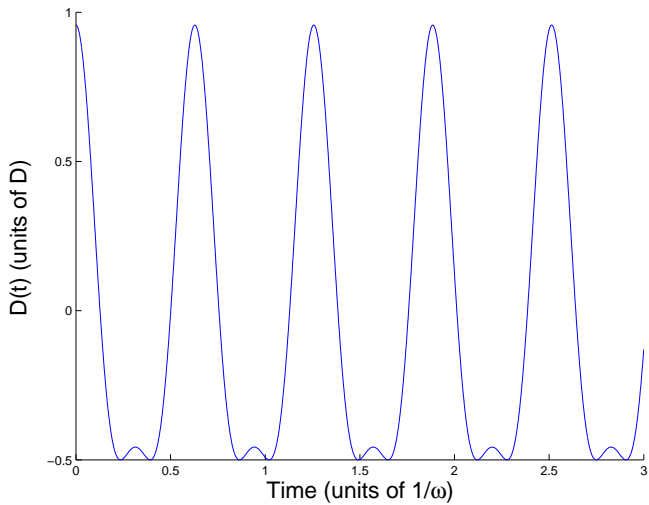}
\end{center}
\caption{The figure shows the modulation of the coupling strength
$D(t)$ as function of time for $\beta = \frac{\pi}{4}$}
 \label{fig:cosplot}
\end{figure}

Now consider the time-dependent case, which models the variation of
coupling strength between homo-nuclear spins under magic angle spinning \cite{Spiess}. The dipolar
interaction strength $D(t)$ during magic angle spinning varies in time as
$D(t)=D \frac{3\cos^2(\theta(t))-1}{2}$, where $\theta(t)$ is the angle internuclear vector makes with the
$B_0$ field. The angle $\theta(t)$ changes as the sample is being rotated around an axis making an angle
$\theta_M = \tan^{-1}(\sqrt{2})$ with the $B_0$ field. Let $\beta$ denote the angle internuclear axis makes
with the magic angle axis. Then we can express $\theta(t)$ as

$$ \cos(\theta(t))=\cos(\beta)\cos(\theta_M)+\sin(\beta)\cos(\omega t )\sin(\theta_M), $$
where $\omega$ is the spinning frequency. $D(t)$ is then a periodic function. We choose
$\beta= \frac{\pi}{4}$ and plot modulation of $D(t)$ in figure ~\ref{fig:cosplot}.
Each period of $D(t)$ can be divided into two parts, $\{D(t)\leq 0\}\bigcup
\{D(t)> 0\}$. Let $S_1$, $S_2$ denote the area of these two parts
respectively, i.e., $S_1=-\int_{\{D(t)\leq 0\}}D(t)\ud t$,
$S_2=\int_{\{D(t)> 0\}}D(t)\ud t$. We find that $S_1=S_2=\frac{1.4922}{\omega}D$.

The canonical parameters for $H_d(t)$ are
\begin{displaymath}
\left\{ \begin{array}{ll}
 D(t)(2,1,-1) & \textrm{for }D(t)\geq0\\
-D(t)(2,1,1) & \textrm{for }D(t)<0\\
\end{array} \right.
\end{displaymath}
i.e., $(2|D(t)|,|D(t)|,-D(t))$. Using theorem (\ref{thm:time}), we
get the minimum time to generate $U$ is the smallest $T$ that
satisfies
$$\frac{3\pi}{4}\leq \int_{0}^{T}3|D(t)|-D(t) \ud t$$
when $\omega>>D$, $\int_{0}^{T}3|D(t)|-D(t) \ud t$ is
approximately $n(2S_1+4S_2)$, where $n$ is the number of periods
of $D(t)$ within time T, so the minimum $n=\lceil
\frac{3\pi}{4(2S_1+4S_2)}\rceil=\lceil \frac{0.2632\omega}{D}
\rceil$ and the minimum time $T$ is approximately $2 \pi
\frac{n}{\omega}$. The pulse sequence prepares effective
Hamiltonians
$(-\sigma_x\otimes\sigma_x+\sigma_y\otimes\sigma_y+2\sigma_z\otimes\sigma_z)$,
$(-2\sigma_x\otimes\sigma_x+\sigma_y\otimes\sigma_y+\sigma_z\otimes\sigma_z)$
and
$(-\sigma_x\otimes\sigma_x+2\sigma_y\otimes\sigma_y+\sigma_z\otimes\sigma_z)$
for $n/3$ periods each, in the part of the period when $D(t) >0$.
Similarly, we prepare effective Hamiltonians
$(-\sigma_x\otimes\sigma_x-
\sigma_y\otimes\sigma_y+2\sigma_z\otimes\sigma_z)$,
$(\sigma_x\otimes\sigma_x+ 2 \sigma_y\otimes\sigma_y
-\sigma_z\otimes\sigma_z)$ and $(-2 \sigma_x\otimes\sigma_x+
\sigma_y\otimes\sigma_y+ \sigma_z\otimes\sigma_z)$ for $n/3$
periods each, in the part of the period when $D(t) <0$. As before,
we apply a local rotation
$e^{-i\frac{\pi}{2}\sigma_x\otimes\sigma_x}$ in the end.

\section{Conclusion}

In this paper, we studied the problem of time-optimal synthesis of
a unitary transformation for coupled qubits under non-stationary
interactions. Under the assumption that local unitary
transformations can be synthesized arbitrarily fast, we
characterized the time optimal trajectories and the minimal time
to prepare a general two qubit rotation under general time varying
coupling tensor. These results generalize the results presented in
~\cite{Khaneja:01,Hammerer,Vidal} for stationary coupling
Hamiltonians to the non-stationary case. The problem considered in
this paper was motivated by design of time optimal pulse sequences
for controlling coupled spin dynamics in solid state NMR
spectroscopy, where couplings between spins are modulated in time
due to magic angle spinning. The results presented here are of
fundamental interest and may find applications in some
implementations of quantum information processing.





\end{document}